\documentclass[aps, prl, 10pt, twocolumn, longbibliography]{revtex4-1}

\usepackage{textcase}
\usepackage{natbib}
\usepackage{braket}
\usepackage{amsmath}
\usepackage{graphicx}
\usepackage{hyperref}
\usepackage{amssymb}
\usepackage[utf8]{inputenc}

\setcitestyle{square}

\begin{document}
\title{Heating-Induced Long-Range $\eta$-Pairing in the Hubbard Model}

\author{J. Tindall$^{1}$, B. Bu\v ca$^{1}$, J. R. Coulthard$^{1}$, and D. Jaksch$^{1,2}$}
\affiliation{$^1$Clarendon Laboratory, University of Oxford, Parks Road, Oxford OX1 3PU, United Kingdom} 
\affiliation{$^2$Centre for Quantum Technologies, National University of Singapore, 3 Science Drive 2, Singapore 117543}

\date{\today}

\setlength{\parskip}{0pt}
\begin{abstract}
We show how, upon heating the spin degrees of freedom of the Hubbard model to infinite temperature, the symmetries of the system allow the creation of steady states with
long-range correlations between $\eta$-pairs. We induce this heating with either dissipation or periodic driving and evolve the system towards a non-equilibrium steady state, a process which melts all spin order in the system. The steady state is identical in both cases and displays distance-invariant off-diagonal $\eta$ correlations. These correlations were first recognised in the superconducting eigenstates described in C. N. Yang's seminal paper \{\href{https://link.aps.org/doi/10.1103/PhysRevLett.63.2144}{Physical Review Letters, \textbf{63}, 2144 (1989)}\}, which are a subset of our steady states. We show that our results are a consequence of symmetry properties and entirely independent of the microscopic details of the model and the heating mechanism.
\end{abstract}
\maketitle

\textit{Introduction} - 
Driving and dissipation have recently emerged as transformative tools for dynamically evolving quantum systems into non-equilibrium phases with desirable properties \cite{QSE, OpticalMod3, NonEq}. In the context of strongly-correlated many-body systems these tools can drastically alter their microscopic behaviour and manifest a variety of collective and cooperative phenomena at the macroscopic level. 
\par Controlled dissipation, for example, has been shown to be a versatile resource for quantum information and simulation purposes \cite{Stabilization, DissLocExp}. Recent proposals show how, in ultracold atomic systems, Markovian baths which act quasi-locally can drive the system towards pure steady states with exotic properties such as superfluidity \cite{QSE2, QSE3}. These schemes, however, rely on precise engineering of the Lindblad jump-operators in order to target specific states and avoid the system heating up to a generic thermal ensemble.
\par Meanwhile, significant interest has been generated by the superconducting-like states that have been induced in a variety of materials by transient excitation of the vibrational degrees of freedom (d.o.f.) using terahertz laser pulses \cite{OpticalMod1, OpticalMod2}. These optically driven systems are often modelled via Floquet driving - where the Hamiltonian is subject to a time-dependent periodic field \cite{Coulthard, PhononSC1, PhononSC2}. Through careful choice of the driving parameters the effective Hamiltonian can be modified, on comparatively short timescales, to one which favours superconductivity and so the system may transiently reach a superconducting prethermal state. However, Floquet heating \cite{FloquetHeating1, FloquetHeating2, FloquetHeating3} means that in most cases these systems continue to absorb energy from the driving field, heating them up and eventually destroying any superconducting order present.
\par In this Letter we show that, counter-intuitively, the interplay between symmetry and heating can actually lead to the formation of steady states with coherent, long-range correlations. This heating can be achieved either with dephasing or with periodic driving and, as the generation of these correlations is based on symmetry, the microscopic details of the heating mechanism are unimportant. We believe this could mark an important step in explaining why such states are being observed in experiments \cite{OpticalMod1, OpticalMod2} where the stringent conditions required in previous numerical work \cite{Coulthard, Kaneko, PhononSC1, PhononSC2} are not fulfilled.
\par In the aforementioned steady states we demonstrate how the formation of these long-range correlations results from the melting of all order in the complementary symmetry sector where the heating is applied. The competition between different types of order is a mechanism considered to underpin the formation of transient superconductivity \cite{Science, Sentef} and thus we focus our work within this context.
\par Specifically, we apply driving or dephasing induced heating to the spin d.o.f. of bi-partite $D$-dimensional realisations of the Hubbard model, melting any order in this sector and reach robust mixed states with completely uniform long-range correlations between $\eta$-pairs. This pairing is known to provide a possible mechanism for superconductivity \cite{Yang2, EtaSC1, EtaSC2, Sentef}. Compared to previous proposals, which excite $\eta$-paired states through carefully tuned driving or dissipation \cite{QSE2, Kaneko}, our results are based on symmetry arguments. This means that the engineered steady states are independent of the model parameters and, for arbitrary initial states, guaranteed to have completely uniform $\eta$-correlations. In the case of dephasing we prove the results of our simulations by block-diagonalising the Liouvillian superoperator, giving an explicit parametric form for the steady states.

 %The Hubbard model is a celebrated quantum lattice model, originally proposed for studying electron behaviour in solids, which has found itself as a potential model for high temperature superconductivity \cite{HubbardSC}, realisable in ultracold atom experiments \cite{HubbardExperiment} and, in one dimension, solvable using the Bethe ansatz \cite{HubbardBethe, HubbardModel, Shastry}. 
\textit{Model} - 
The Hubbard model, in second quantised form, reads
\begin{align}
H = -\tau\sum_{\langle ij \rangle, \sigma}(c^{\dagger}_{\sigma, i}c_{\sigma, j} + {\rm h.c}) + U\sum_{i}n_{\uparrow, i}n_{\downarrow, i},
\label{HubbardHam}
\end{align} 
where $c_{\sigma, i}^{\dagger}$ and its adjoint are the usual creation and annihilation operators for a fermion of spin $\sigma \in \{\uparrow, \downarrow\}$ on site $i$. Additionally, $n_{\sigma, i}$ is the number operator for a particle of spin $\sigma$ on site $i$ and $\tau$ and $U$ play the role of kinetic and interaction energy scales respectively. In this work we set $\hbar = 1$. 
%\par The Hubbard model has long been studied for its potential relevance to high temperature superconductivity. For example, in the limit of strong interactions, one can project out states which contain double occupancies, resulting in the $t$-$J$ model, a model of strongly correlated electrons which interact via an attractive super-exchange interaction.  Importantly, for this work, it has recently been demonstrated that optical modulation at a frequency resonant with $U$ can excite eta-paired states \cite{Kaneko}.
\par 
The Hamiltonian in Eq. (\ref{HubbardHam}) has a rich structure due to the two SU(2) symmetries it possesses \cite{HubbardModel}. The first of these is the `spin' symmetry which relates to spinful particles (singlons) $\sigma \in \{\uparrow, \downarrow\}$. The second, often referred to as `$\eta$-symmetry', is central to this letter and relates to spinless quasi-particles (doublons and holons) $\sigma \in \{\uparrow\downarrow, {\rm vac}\}$. It can be interpreted as a type of particle-hole symmetry. This $\eta$-symmetry is revealed by introducing the associated operators
\begin{align}
\eta^{+} &= \sum_{i}\eta^{+}_{i} = \sum_{i}(-1)^{i}c^{\dagger}_{\uparrow, i}c^{\dagger}_{\downarrow, i}, \notag \\ \eta^{-} &= \sum_{i}\eta^{-}_{i} = \sum_{i}(-1)^{i}c_{\downarrow, i}c_{\uparrow, i}, \notag \\ \quad \eta^{z} &= \sum_{i}\eta^{z}_{i} = \sum_{i}\frac{1}{2}(n_{i, \uparrow} + n_{i, \downarrow} - 1),
\label{Eq: Eta-Operators}
\end{align}
with $\eta^{+}_{i}$ ($\eta^{-}_{i}$) describing the creation (annihilation) of a doublon on site $i$ with an alternating site-dependent phase. The operators in Eq. (\ref{Eq: Eta-Operators}) fulfill the relations $[H, \eta^{\pm}] = 0$ and $[H, \eta^{+}\eta^{-}] = [H, \eta^{z}] = 0$ and commute with all the generators of the spin symmetry.
\par The presence of $\eta$-pairing superconductivity in the Hubbard model is a phenomenon first recognised by C. N. Yang in his seminal paper \cite{Yang1}. There it was proved that the pure states $\ket{\psi} \propto (\eta^{+})^{N}\ket{{\rm vac}}$ are eigenstates of $H$ and possess off-diagonal long-range order (ODLRO) in the form of doublon-doublon correlations
\begin{equation}
\label{ODLRO}
{\rm Tr}(\rho \eta^{+}_{i}\eta^{-}_{j}) = {\rm const.} \quad \forall i,j \quad i \neq j,
\end{equation} 
where $\rho = \ket{\psi}\bra{\psi}$. This relation provides a possible definition of superconductivity as a finite value of this quantity can be shown to imply both the Meissner effect and flux quantisation \cite{Yang2, Meissner1, Meissner2}. These states, however, are excited states of $H$ and the long-range order they possess is not usually seen in physical states (ground states, thermal states etc.) of the model due to destructive interference from the short-range coherences involving spinful particles. By driving the Hubbard model in the spin basis these are destroyed and we can consistently engineer states with long-range uniform correlations in $\langle \eta^{+}_{i}\eta^{-}_{j} \rangle$.  
\par As our primary mechanism for achieving this we consider the Hubbard model immersed in an environment which induces local dephasing in the spin basis. This model is motivated by the spin fluctuation theory of superconductivity, where electrons pair due to their scattering on spin fluctuations \cite{SpinFluctuations, SpinFluctuationSC1, SpinFluctuationSC2}. Our dephasing mimics these scattering events.
% and recent experiments have demonstrated that spin-photon interactions can be induced by placing spin-electron ensembles within Magnonic cavities \cite{Magnonic2, Magnonic}. 
\par Despite the `toy' nature of our model the Hubbard Hamiltonian can be accurately realised by loading ultracold fermionic atoms into optical lattices \cite{HubbardExperiment, Jaksch, Maciej}. These quantum simulators offer precise experimental control over the microscopic details of the system. In the Supplemental Material (SM) \cite{Note1} we show how dephasing can occur by immersing the lattice into a homogeneous Bose-Einstein condensate \cite{QME1, QME2}. If the interactions are tuned with Feshbach resonances \cite{FeshbachOpt, FeshbachMag} so that the scattering amplitudes between the two fermionic spin states and the bosons are equal in magnitude and opposite in sign, then dephasing will occur solely in the spin sector.
%We also suggest that the dephasing could be achieved by the immersion of the lattice into a bath of atoms which have a spin-spin interaction with the hyperfine levels of the fermionic atoms \cite{SpinDephasing}.

\textit{Results} - We couple the Hubbard Hamiltonian in Eq. (\ref{HubbardHam}) to an environment which applies spin dephasing on each site of an $M$-site lattice. The ensuing dynamics is modelled, under the Markov approximation, via the Lindblad master equation
\begin{align}
\frac{\partial \rho}{\partial t} = \mathcal{L}\rho = -i[H, \rho] + \gamma \sum_{j}&\big(L_{j}\rho L_{j}^{\dagger} - \frac{1}{2}\{L^{\dagger}_{j}L_{j}, \rho\}\big), \notag \\
L_{j} &= s^{z}_{j} = n_{\uparrow, j} - n_{\downarrow, j},
\label{Master Equation}
\end{align}
with Lindblad operators $s^{z}_{j}$ on each site $j$. Because these operators are restricted to the spin SU(2) symmetry sector we obtain $[L_{j}, \eta^{\pm}] = [L_{j}, \eta^{z}] = 0 \ \forall j$. 
\par For Eq. (\ref{Master Equation}) any operator $A$ which satisfies 
\begin{equation}
[H, A] = [L_{j}, A] = [L_{j}^{\dagger}, A] = 0 \quad \forall j,
\label{Eq:Commutators}
\end{equation}  
is a null eigenvector of  $\mathcal{L}$ and the adjoint map $\mathcal{L}^{\dagger}$, which means the associated observable $\langle A \rangle$ is conserved. Thus, the map in Eq. (\ref{Master Equation}) conserves $\langle \eta^{+}\eta^{-} \rangle$, $\langle \eta^{z} \rangle$ and $ \langle S^{z} \rangle = \sum_{i} \langle s^{z}_{i} \rangle$. Through a power series expansion, any appropriately trace normalised function $f(\eta^{+}\eta^{-}, \eta^{z},S^{z})$ of these operators is a steady state of $\mathcal{L}$, i.e. if $\rho_{ss} = f(\eta^{+}\eta^{-}, \eta^{z},S^{z})$ then $\mathcal{L}\rho_{ss} = 0$. In the SM  \cite{Note1} we show that steady states of this form have ${\rm Tr}(\rho_{ss} \eta^{+}_{i}\eta^{-}_{j}) = {\rm const} \ \forall i,j$ (see also \cite{DarkHamiltonians}). We also show how this result of heating-induced long-range order can be extended to arbitrary models with multiple SU(2) symmetries. We emphasize that the expected steady state $\rho_{ss}$ is realisable for any bi-partite $D$ dimensional realisation of the spin-dephased Hubbard model.

\par Here, we focus on 1D lattice realisations due to their numerical tractability. We perform calculations where the system is initialised in a given state and then evolved by solving the time-dependent master equation in Eq. (\ref{Master Equation}) and quenching $U$, providing a direct comparison with the dynamics of the closed system. In the open system the value of $U$ quenched to solely sets the time-scale of relaxation and long-range order will always appear. We use a quantum trajectories approach \cite{Trajectories} to solving Eq. (\ref{Master Equation}) combined with Density Matrix Renormalisation Group \cite{DMRG} and Time Evolving Block Decimation \cite{TEBD} algorithms for finding the ground state and evolving the wavefunction of the system respectively. These simulations were performed with the Tensor Network Theory library \cite{TNT}. Further numerical details and analysis can be found in the SM  \cite{Note1}.

\begin{figure}[t]
\includegraphics[width = \columnwidth]{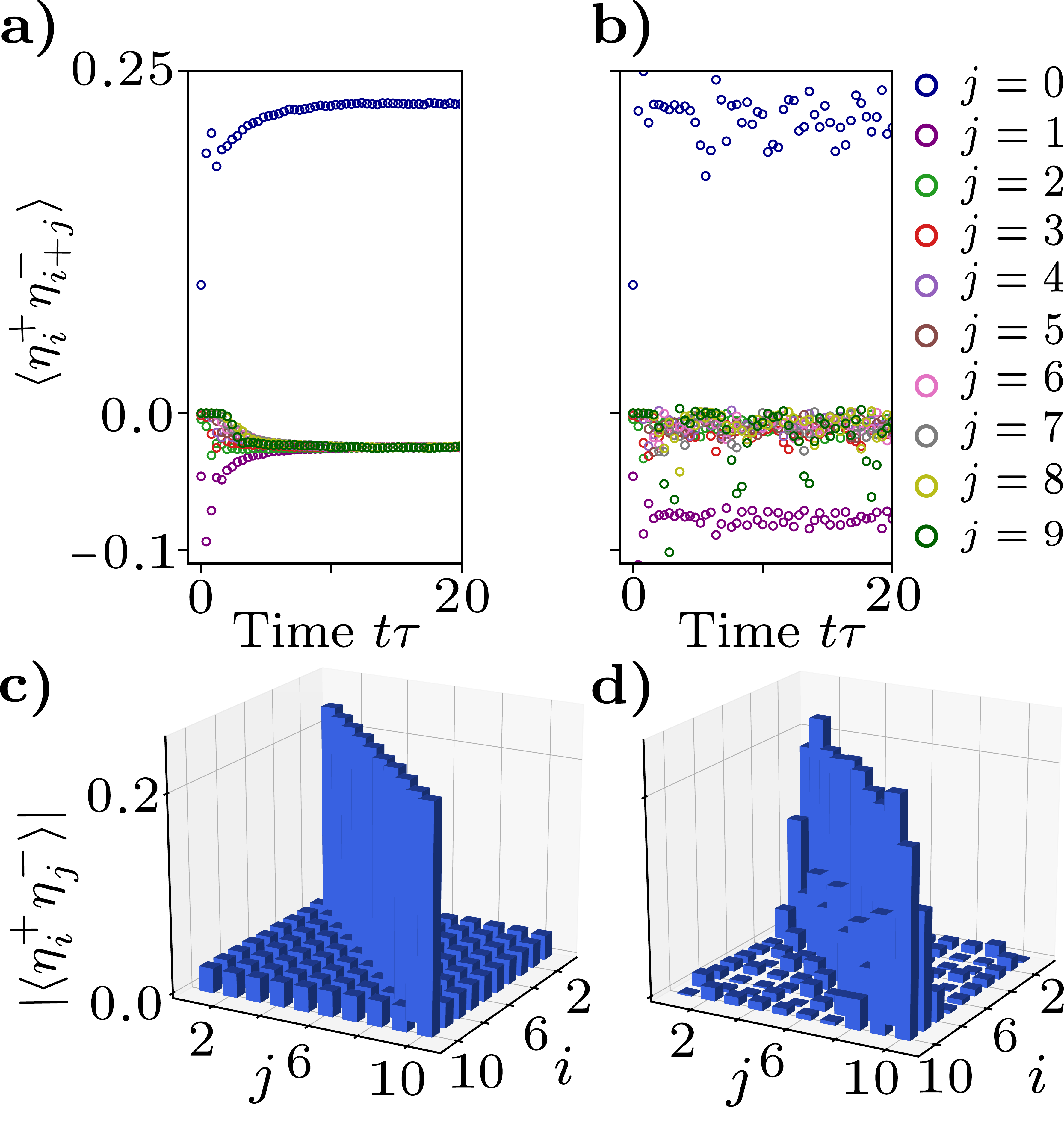}
\caption{(a) and (b) Evolution of the correlator $\braket{\eta^{+}_{i}\eta^{-}_{i + j}}$, averaged over all sites separated by a distance $j$, for the $10$ site Hubbard model at half-filling and $\langle S^{z} \rangle = 0$. The system is initialised in the ground state of $H$ for $U = 4.0 \tau$ and then evolved with a system-environment coupling of $\gamma = 2.0 \tau$ for (a) and $\gamma = 0.0$ for (b). In both quenches the interaction strength is changed to $U = \tau$. (c-d) Matrix of correlations of $|\langle \eta^{+}_{i}\eta^{-}_{j} \rangle|$ associated with the density matrix at $t \tau = 20.0$ for the simulations (a-b) respectively.}
\label{Fig:ODLRO}
\end{figure}

\par In Fig.~\ref{Fig:ODLRO}. we plot, for various distances $j$, the averaged quantity $\braket{\eta^{+}_{i}\eta^{-}_{i + j}}$ over time for a quench from the ground state of the Hubbard Hamiltonian. We also include the matrix of correlations of $\langle \eta^{+}_{i} \eta^{-}_{j} \rangle$  for the density matrix following the quench. For comparison, we show the case where the system is not coupled to the environment, $\gamma = 0$. When the environment is present there is a relaxation of the system to a steady state which possesses the order expressed in Eq. (\ref{ODLRO}); the value of $\langle \eta^{+}_{i} \eta^{-}_{j} \rangle$ is finite and constant for all $i \neq j$. We observe that this is facilitated by a decrease in the short-range $\eta$-pairing correlations in order to allow the long-range ones to increase: the non-equilibrium dynamics involve a `spreading' of the $\eta$-correlations over all length scales of the system which is necessary due to the conservation of $\eta^{+}\eta^{-}$. Plots of the corresponding doublon momentum distribution are discussed in the SM \cite{Note1}.
\par We also show how this ODLRO is observable even under perturbative and unwanted dephasing in the model. We do this by introducing `charge-dephasing' using jump operators that don't solely act in the spin basis, breaking the strong symmetry relations and causing the steady state of the Liouvillian to contain no coherences. Nonetheless, we set the strength of the charge-dephasing to values around $1 \%$ of the spin-dephasing and find there exists an intermediate time-scale where the results of Fig.~\ref{Fig:ODLRO} can be observed. We also find that the window in which the uniform long-range correlations exist can be directly controlled by the ratio of the couplings for the spin dephasing and the charge dephasing: the dissipative evolution induced by the charge-dephasing becomes increasingly frozen out as this ratio increases. In particular, our data suggests that the length of this window can be extended by increasing the coupling of the spin dephasing whilst the other parameters in the model remain constant. 
\par To obtain a more intuitive idea of the nature of the steady state $\rho_{ss} = f(\eta^{+}\eta^{-}, \eta^{z},S^{z})$ we consider a physically motivated parametrisation of the steady state function
\begin{align}
\rho_{ss} \propto \exp(\mu_{1}\eta^{+}\eta^{-} + \mu_{2}N_{\uparrow} + \mu_{3}N_{\downarrow}),
\label{Effective Thermal State}
\end{align}  
where we have exchanged $\eta^{z}$ and $S^{z}$ with $N_{\uparrow} = \sum_{i}n_{\uparrow, i}$ and $N_{\downarrow} = \sum_{i}n_{\downarrow, i}$ as they can be expressed as linear combinations of these operators. Calculations on small lattices show that the parametrisation in Eq. (\ref{Effective Thermal State}) captures all relevant observables when compared to numerical simulations which reach the long-time limit in Eq. (\ref{Master Equation}). Equation (\ref{Effective Thermal State}) describes a generalized grand-canonical-like equilibrium state (GCE) with the Lagrange multipliers $\mu_{1}, \mu_{2}, \mu_{3}$ associated to each of the conserved quantities. Notably, the Lagrange multiplier for the Hamiltonian is $0$, i.e. at infinite temperature. As is typical of states in this form the Lagrange multipliers are independent of the quench parameters and are governed solely by the values of the conserved quantities $\langle \eta^{+}\eta^{-}\rangle $, $\langle N_{\uparrow} \rangle$ and $\langle N_{\downarrow} \rangle$ associated with the initial state. The existence of a `quadratic' charge ($\eta^{+}\eta^{-}$) in the GCE state is crucial for the presence of non-trivial long-range order and sets it apart from Generalised Gibbs Ensemble states \cite{ExactGGE, GGE1, GGE2, Jordi}.

\begin{figure}[t]
\includegraphics[width = \columnwidth]{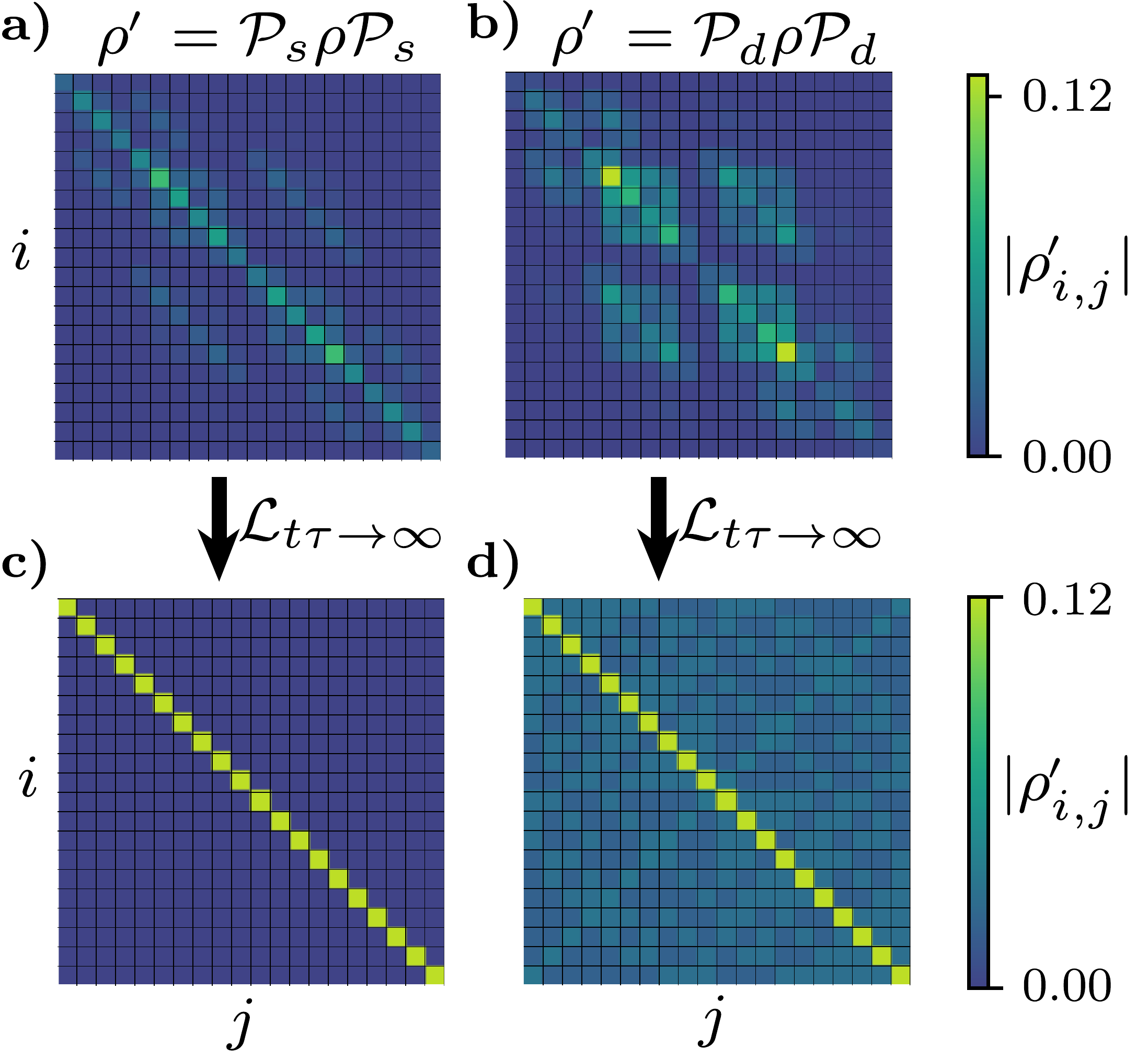}
\caption{Projections, $\rho' = \mathcal{P}_{s, d}\rho\mathcal{P}_{s, d}$, of the density matrix in the spin ($s$) and $\eta$ ($d$) sectors of the $M = 6$ Hubbard model. The system is at symmetric half-filling $\langle N_{\uparrow} \rangle = \langle N_{\downarrow} \rangle = M/2$ and the projections have been renormalised (${\rm Tr}(\rho') = 1$). The color
indicates the magnitude of the matrix elements. Following the projection, the indices $i$ and $j$ run over the remaining basis states in lexicographic order when they are converted to binary strings where $\uparrow = 1$, $\downarrow = 0$ and $\uparrow\downarrow = 1$, $\rm vac = 0$ for the projections in $s$ and $d$ respectively. As an example, for plots (a) and (c), when $i = 1$ this corresponds to the basis vector $\ket{\uparrow \uparrow \uparrow \downarrow \downarrow \downarrow} = \ket{111000}$ and when $i = 20$: $\ket{\downarrow \downarrow \downarrow \uparrow \uparrow \uparrow} = \ket{000111}$.  (a) and (b) Projections for the thermal state of the $U = 2.5\tau$ Hubbard model: $\rho \propto \exp(-\beta H)$ with $\beta = 0.8/\tau$. (c) and (d) Projections following spin-dephasing of this initial state by the map in Eq. (\ref{Master Equation}) in the long-time limit with quench parameters $U = \tau$ and $\gamma = 2.0\tau$.}
\label{Fig:Projections}
\end{figure}

\par To emphasize the properties of the stationary state in Eq. (\ref{Effective Thermal State}) we plot, in Fig.~\ref{Fig:Projections}., the projection of the density matrix in the spin ($s$) and $\eta$ ($d$) d.o.f. $\mathcal{P}_{s, d}\rho\mathcal{P}_{s, d}$. Explicitly, these projectors remove any basis vectors which contain, respectively, doublons ($\sigma \in \{\uparrow\downarrow, \rm vac \}$) and singlons ($\sigma \in \{\uparrow, \downarrow\}$) on any lattice sites. We plot these projections following a quench from a thermal state of the Hubbard model. Initially (Figs. \ref{Fig:Projections}a. and \ref{Fig:Projections}b.), the system contains coherences in both these sectors which decay with distance - additionally (not pictured) the density matrix contains coherences between doublons and singlons. Following spin dephasing in the long-time limit any coherences involving singlons are destroyed resulting in an infinite temperature ensemble in the spin symmetry sector (Fig.~\ref{Fig:Projections}c.). Only coherences involving doublons and holons remain (Fig.~\ref{Fig:Projections}d). These are completely distance-invariant, as described by the parametrisation in Eq. (\ref{Effective Thermal State}).

\begin{figure}[t]
\includegraphics[width = \columnwidth]{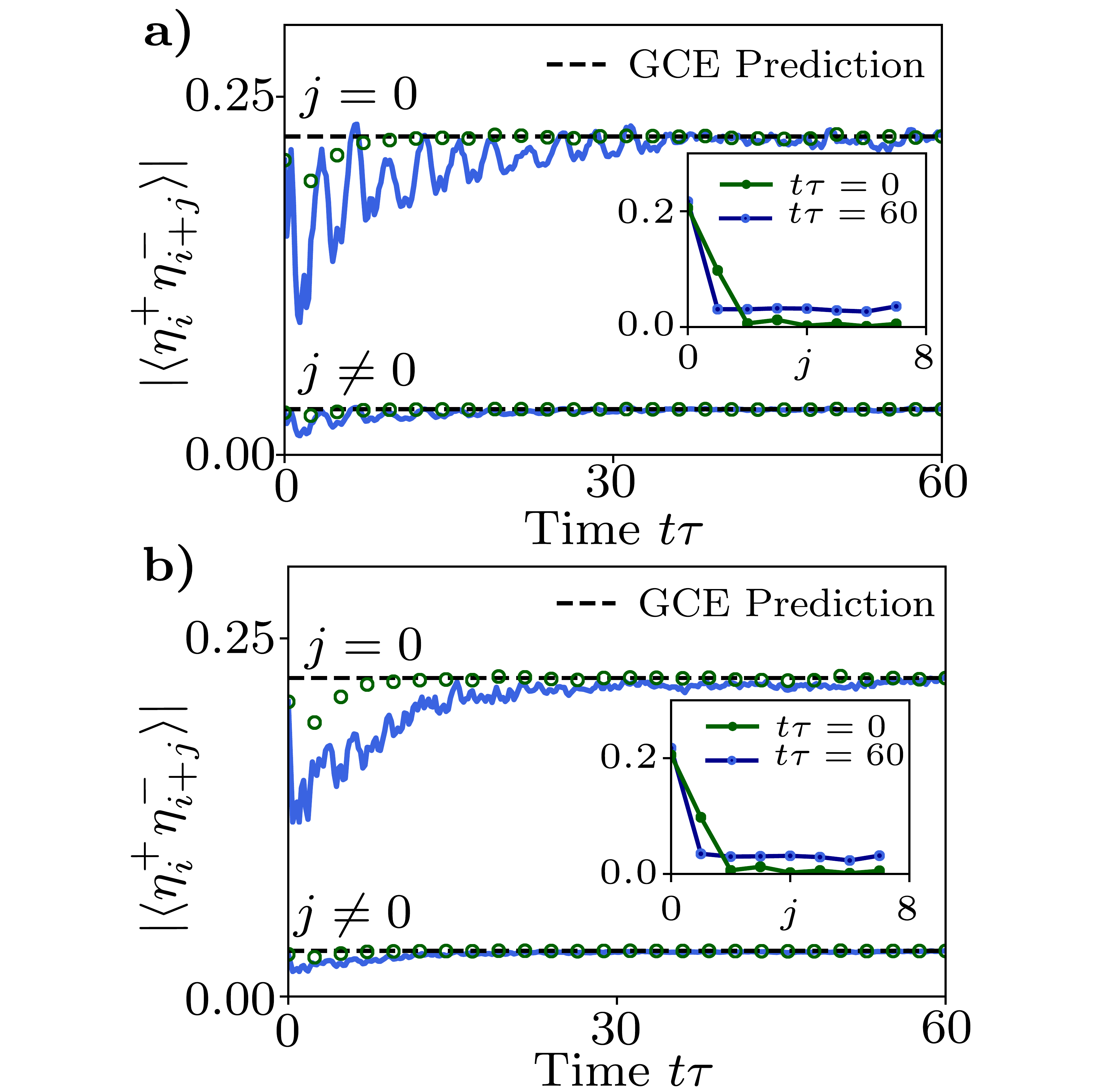}
\caption{Dynamics of the averaged $\eta$ correlations $|\langle \eta^{+}_{i}\eta^{-}_{i + j} \rangle|$ following a quench from the $U = \tau$ ground state of the $M=8$ half-filled Hubbard model with $\langle S^{z} \rangle = 0$. The blue curves reflect a quench with the periodically driven Hamiltonian in Eq. (\ref{Eq:Driven Hamiltonian}) where $U = 6.0\tau$ and $\Omega = \tau$. The green markers are for a quench under the dephasing map in Eq. (\ref{Master Equation}) with $U = \tau$ and $\gamma = 2.0\tau$. The dashed lines indicate the prediction from the grand-canonical ensemble in Eq. (\ref{Effective Thermal State}). Insets show the dependence of the correlations on distance, extracted for the blue curve at times $t\tau = 0.0$ and $t\tau = 60.0$. (a) $V = 2.0U$ and $f(i) = i$. (b) $V = 4.0U$ and $f(i) \in \rm Rand[0,2.0]$, a uniform random number on the specified interval.}
\label{Fig:Floquet Heating1}
\end{figure}

\par The strong symmetries \cite{Prosen} $\langle \eta^{+}\eta^{-}\rangle $, $\langle N_{\uparrow} \rangle$ and $\langle N_{\downarrow} \rangle$ can be used to block-diagonalise the Liouvillian and hence the degeneracy of $\rho_{ss}$ is determined by the distinct combinations of the eigenvalues of these operators. The steady states are the states with maximum $\eta$-order in each block, quantified by the amplitude of the uniform off-diagonal correlations $|\langle \eta^{+}_{i}\eta^{-}_{j}\rangle|, \ \ i\neq j$. Notably, Yang's states $\ket{\psi} \propto (\eta^{+})^{N}\ket{{\rm vac}}$ form a subset of these steady states. They are the only pure steady states as they exist in the decoherence free subspace \cite{DFS1} of Eq. (\ref{Master Equation}).
\par The infinite temperature ensemble pictured in Fig.~\ref{Fig:Projections}c. shows that the dephasing has continuously pumped energy into the spin d.o.f. of the lattice until it saturates and reaches infinite temperature. An alternative method of achieving this is through Floquet Heating. To show this we consider the closed Hubbard model under periodic driving in the form of a time-dependent inhomogeneous magnetic field
\begin{equation}
H(t) = H +  B(t)\sum_{i = 1}^{L}f(i)s^{z}_{i},
\label{Eq:Driven Hamiltonian}
\end{equation}
with $B(t) = V\cos(\Omega t)$ and $f(i)$ describing the inhomogeneity. In the long-time limit the driving is expected to thermalise the system to infinite temperature \cite{FloquetHeating1, FloquetHeating2}. However, because in this case the driving term commutes with all the generators of the $\eta$-symmetry we expect heating to occur only within the spin sector.

%\begin{figure}[t]
%\includegraphics[width = \columnwidth]{Images/Figure4.pdf}
%\caption{(a) Dynamics of the the $\eta$ correlations $\langle \eta^{+}_{i}\eta^{-}_{i + j} \rangle$ following a quench from the pulse irradiated ground state of the half-filled $L = 10$ Hubbard model. Only even distances are shown for clarity. (b) Decay of the same correlations with distance $j$ for the pulse irradiated state ($t \tau = 0.0$) and the same state following the quench in (a).}
%\label{Fig:Floquet Heating2}
%\end{figure}

\par In Fig. \ref{Fig:Floquet Heating1}. we show how, by time-evolving an initial state under $H(t)$, we realise long-time dynamics identical to that induced by the spin-dephasing. By resonantly driving the system, $U = n\Omega \ n \in \mathbb{Z}^{+}$, the system thermalises quickly \cite{FloquetHeating3} and the observables are in good agreement with both the grand-canonical description of (\ref{Effective Thermal State}) and long-time simulations of the map in Eq. (\ref{Master Equation}). There is completely uniform off-diagonal long-range order in $\eta$-pairs (Fig.~\ref{Fig:Floquet Heating1}. Insets). 
\par In Fig. \ref{Fig:Floquet Heating1}a. we choose a linear magnetic field $f(i) = i$, whilst in Fig. \ref{Fig:Floquet Heating1}b. we choose a disordered field. In both cases the long-time dynamics are the same and this emphasizes that the choice of driving parameters and inhomogeneity is arbitrary - they only affect the time-scale on which the system relaxes.

%\par In Figure. \ref{Fig:Floquet Heating1}b), we combine this driving with an initial state that has been pre-prepared by optical modulation. This technique is based on the work by Kaneko et al \cite{Kaneko} which uses a Guassian shaped pulse to irradiate the lattice and increase the $\eta$-correlations. This effect, however, is only local and so it does not guarantee uniform long-range correlations in the wavefunction. Following Floquet heating in the spin sector we ensure these uniform long-range correlations - which are larger compared to heating the ground state of the Hubbard model directly.  

\textit{Conclusion} - We have demonstrated that long-range $\eta$-pairing can be created and protected within the Hubbard model by directly heating the spinful d.o.f. to infinite temperature. This destroys any coherences involving spinful particles which, in turn, creates uniform long-range correlations between the $\eta$ quasi-particles in the lattice.
\par Recent work has demonstrated that, with a judicious choice of parameters, applying a Gaussian pulse of a specific duration to the Mott-Insulating phase of the Hubbard model can excite $\eta$-pairs \cite{Kaneko}. This driving does not commute with the generators of the $\eta$-symmetry and hence uniform long-range correlations are not guaranteed. This does, however, open up the possibility of applying our heating scheme in conjuction with a similar driving process which can enhance the $\eta$-correlations in the system. Our heating would then continually spread these over all length-scales, creating a long-lived superconducting state.   
\par We also anticipate further work using an alternate heating mechanism within the Hubbard model, or $t$-$J$ model, which protects singlets and destroys other coherences. This could enhance superexchange pairing and induce superconductivity through nearest-neighbour singlet pairing \cite{Coulthard, Singlets1, Singlets2}. 
\par Finally, we emphasize the results of this work are due to the multiple ${\rm SU}(2)$ symmetries of the Hubbard model and not its microscopic details. Hence, ODLRO can be realised through dephasing in any model with multiple symmetries (see SM  \cite{Note1}), as examples we suggest multi-band Hubbard models and the Richardson-Gaudin model \cite{Gaudin} which also permit superconducting regimes. 
\\
\begin{acknowledgments}
We would like to acknowledge S.R. Clark for our fruitful conversations and his beneficial input on optical modulation. We also thank C. S\'anchez Mu\~noz, J. Mur-Petit, H. Gao and J. De Nardis for useful discussions. This work has been supported by EPSRC grants No. EP/P009565/1 and EP/K038311/1 and is partially funded by the European Research Council under the European Union’s
Seventh Framework Programme (FP7/2007-2013)/ERC Grant Agreement No. 319286 Q-MAC. We
acknowledge the use of the University of Oxford Advanced Research Computing (ARC) facility in
carrying out this work http://dx.doi.org/10.5281/zenodo.22558.
\end{acknowledgments}

\pagebreak

\section{Supplementary Material To Heating-Induced Long-Range $\eta$-Pairing in the Hubbard Model}
\section{Numerical Procedure}
The master equation we solve in the main text reads
\begin{align}
\label{Master EquationS}
\frac{\partial \rho}{\partial t} = \mathcal{L}\rho = -i[H, \rho] + \gamma \sum_{j =1}^{M}(L_{j}\rho L_{j}^{\dagger} - \frac{1}{2}\{L^{\dagger}_{j}L_{j}, \rho\}), \quad \notag \\
L_{j} = s^{z}_{j} = n_{\uparrow, j} - n_{\downarrow, j},
\end{align}
with 
\begin{align}
H = -\tau\sum_{\langle ij \rangle, \sigma}(c^{\dagger}_{\sigma, i}c_{\sigma, j} + {\rm h.c}) + U\sum_{i = 1}^{M}n_{\uparrow, i}n_{\downarrow, i}.
\label{HubbardHamS}
\end{align} 

For our numerical simulations we perform a stochastic unravelling of this master equation, known as the `quantum trajectories' approach \cite{Trajectories}. We use DMRG (Density Matrix Renormalisation Group) \cite{DMRG} to find the ground state of the Hamiltonian and then evolve this in time using the TEBD (Time Evolving Block Decimation) \cite{TEBD} algorithm. We focus on 1D lattice realisations of the model due to their numerical tractibility. We implemented these algorithms using the TNT (Tensor Network Theory) Library \cite{TNT}. For all figures within the main text we use a bond-dimension of $\chi = 1000$ to ensure the corresponding SVD (Singular Value Decomposition) errors are minimal and the sum of the squares of the discarded singular values in a given time-step does not exceed $\epsilon = 1 \times 10^{-4}$. Increasing the bond dimension has no effect on our results. For all simulations with a finite value of $\gamma$ we perform $N = 2000$ trajectories to ensure convergence of the measured observables to within an uncertainty of $2 \%$. We also checked that lowering the timestep $\delta t$ from the value ($\delta t = 0.01 / \tau$) used in our second-order Trotter decomposition of the propagator has no noticeable effect on our results.
\par Despite using a TEBD approach we are not able to reach particularly large system sizes as the states reached in the long-time limit have both completely long-range entanglement and a significant amount of classical correlations. This can cause the Matrix Product State representation to be highly inefficient \cite{BondDimension} and we find the bond dimensions required for our simulations, at least at the levels of the individual trajectories, is large and prevents us from being able to make effective use of the aforementioned Matrix Product algorithms. 
\par
\begin{figure*}[t!]
\centering
\includegraphics[width = \textwidth]{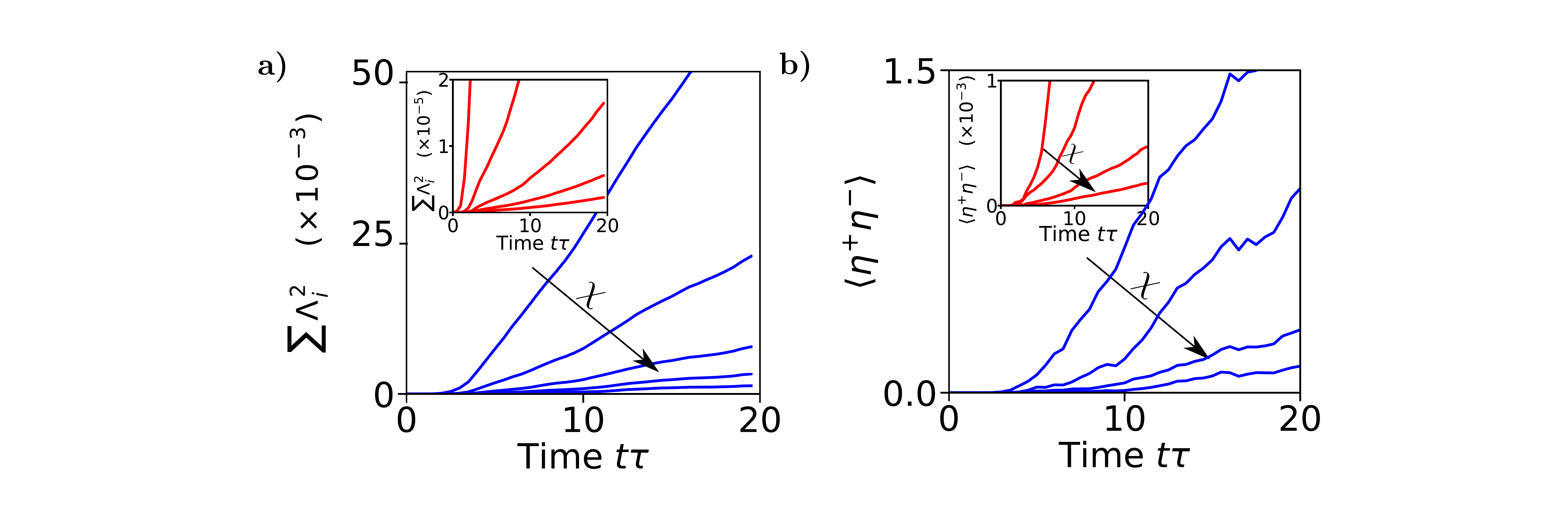}
\caption{Simulation starting from the ground state of the $U = \tau$ Hubbard model for $L = 10$ and symmetric half-filling $\big( \langle N_{\uparrow} \rangle = \langle N_{\downarrow} \rangle = L/2 \big)$. The system is then time-evolved, with the TEBD algorithm, under the Master Equation in Eq. (\ref{Master EquationS}) but with $U = 2.0\tau$. The bond dimensions used are $\chi = 100, 200, ..., 500$ with larger $\chi$ corresponding to the curves closer to the $x$-axis. (a) Cumulative total of the squares of the discarded singular values in the TEBD algorithm, averaged over $10$ trajectories, for the open system with $\gamma = 2.0\tau$. Inset) Cumulative total of the squares of the discarded singular values in the TEBD algorithm for the closed system ($\gamma = 0.0$). (b) Value of $\langle \eta^{+} \eta^{-} \rangle$ during the simulation (which should, for each trajectory, be conserved) for the open system with $\gamma = 2.0\tau$. Inset) Value of, trajectory-averaged, $\langle \eta^{+} \eta^{-} \rangle$ during the simulation for the closed system ($\gamma = 0.0$).}
\label{Fig:SVD}
\end{figure*}

 In Fig. \ref{Fig:SVD} we demonstrate this: the cumulative total of the squares of the discarded singular values (a measure of the accuracy of the simulation) over time for an average trajectory in the open system is several orders of magnitude larger than that of the corresponding closed system. Unless a very large bond dimension is used this results in a critical failure of the numerics to conserve physical symmetries of the system, such as $\langle \eta^{+} \eta^{-} \rangle$ - which should, for each trajectory, always remain $0$ in the example considered.

\section{Using Symmetry Based Dephasing to Induce Distance Invariant Off-Diagonal Correlations}
\label{Sec:Proof}
\par The result of engineering off-diagonal long-range correlations through dissipation is not specific to the Hubbard model nor the choice of Lindblad operators, it is a consequence of the multiple SU(2) symmetries of the Hamiltonian. To illuminate this we define a general Liouvillian map $\Lambda$ in the manner of (\ref{Master EquationS}), with a Hamiltonian $H$ and set of jump operators $\{L_{k}\}$. We consider the situation when $H$ has at least $2$ SU(2) symmetries, i.e. there exists $N \geq 2$ sets of operators $\{J_{i}^{\pm, z}\} \quad i = 1, 2, ..., N$ where $J_{i}^{\pm, z}$ are the $3$ generators for the SU(2) symmetry $i$ of the Hamiltonian
\begin{align}
[H, J_{i}^{z}] &= 0, \quad [H, J_{i}^{\pm}] = \pm \mu_{i} J_{i}^{\pm}, \quad [H, J_{i}^{+}J_{i}^{-}] = 0, \quad \notag \\ [J_{i}^{z}, J_{i'}^{\pm}] &= \pm \delta_{i, i'} J_{i}^{\pm}, \quad [J_{i}^{+}, J_{i'}^{-}] = 2 \delta_{i, i'}J_{i}^{z}, \quad \mu_{i} \in \mathbb{R}.
\label{Eq: Symmetries}
\end{align}
The commutativity between any pairs of generators from two different symmetries is encoded by the Kronecker delta. Any of the operators can be written as a sum of their projectors, via an eigendecomposition, \\ $J_{i}^{+-, z} = \sum_{\alpha = 1}^{D_{i}^{+-, z}}\lambda_{i, \alpha}^{+-, z}\mathcal{P}_{i, \alpha}^{+-, z}$. We have combined $J_{i}^{\pm}$ to form $J_{i}^{+-} = J_{i}^{+}J_{i}^{-}$ due to their mutual eigenspaces.
\par We now induce dephasing on the model in a symmetry protected manner by choosing $\{L_{k}\}$ such that 
\begin{equation}
[L_{k}, J_{j}^{+-,z}] = 0,
\label{Eq: Symmetries2}
\end{equation}
for at least one symmetry $j$ within $i = 1,2, ..., N$. Provided that the map is unital ($\sum_{k}[L_{k}, L^{\dagger}_{k}] = 0$) then the commutation relations in Eq. (\ref{Eq: Symmetries}) result in any projector associated with a generator satisfying (\ref{Eq: Symmetries2}) being a null eigenvector of the Liouvillian map $\Lambda(\mathcal{P}_{j, \alpha}^{+-, z}) = 0$ \cite{Prosen}. The projectors $\mathcal{P}_{j, \alpha}^{+-, z}$ for all the $j$ satisfying (\ref{Eq: Symmetries2}), and those for the operators  $\{C\}$ relating to any remaining conserved quantities \cite{Albert}, fully span the kernel of $\Lambda$. It then follows that linear combinations (or any trace-normalised function which can be expanded as a power series)  of these projectors is a steady state of the map $\Lambda$ \cite{Prosen}. 
\par Let us write the steady states in this form
\begin{equation}
\rho_{ss} \propto \sum_{\beta}\lambda_{\beta}\mathcal{P}_{\beta},
\label{Eq:Steady State}
\end{equation}
where the sum is over all the projectors spanning the kernel of $\Lambda$. The following analysis is also valid if we write the steady state as any power-series expandable function of these projectors. Assuming we have a lattice model we define a permutation operator $P_{i, i'}$ ($P_{i, i'}^{2} = 1$) which exchanges sites $i$ and $i'$. If all the operators, $\{O\} = \{J_{j}^{+-,z}\}\ \bigcup \ \{C\}$, whose projectors are involved in Eq. (\ref{Eq:Steady State}) are invariant under the same $P_{i, i'}$ then
\begin{equation}
P_{i, i'}OP_{i, i'} = O \implies [O, P_{i, i'}] = 0 \implies [\mathcal{P}_{\beta}, P_{i, i'}] = 0 \quad \forall \beta.
\label{Eq:Symm3}
\end{equation}
It immediately follows that
\begin{equation}
[\rho_{ss}, P_{i ,i'}] = 0.
\label{Eq:Symm4}
\end{equation}
Now let us choose one of the symmetries satisfying (\ref{Eq: Symmetries2}) and index it by $k$. The $3$ generators for this symmetry are $J_{k}^{+,-,z}$ and all of their projectors are contained in Eq. (\ref{Eq:Steady State}). We assume they can be written as a sum of purely local operators on the $M$ site lattice: $J_{k}^{+,-,z} = \sum_{l = 1}^{M}J_{k, l}^{+,-,z}$. Then we  define the off-diagonal expectation value
\begin{equation}
\langle J^{+}_{k, l}J^{-}_{k, m} \rangle = {\rm Tr}(\rho_{ss}J^{+}_{k, l}J^{-}_{k, m}) \quad l \neq m.
\end{equation}
We now prove this is invariant under the choice of $l$ and $m$ by using the resolution of identity with the permutation operator $P_{j ,j'}$ as well as (\ref{Eq:Symm4}) 
\begin{align}
&{\rm Tr}(\rho_{ss}J^{+}_{k, l}J^{-}_{k, m}) = {\rm Tr}(\rho_{ss}P_{l, l'}^{2}J^{+}_{k, l}P^{2}_{m, m'}J^{-}_{k, m}) = \notag \\ &{\rm Tr}(\rho_{ss}P_{l, l'}J^{+}_{k, l}P_{l, l'}P_{m, m'}J^{-}_{k, m}P_{m, m'}) = \notag \\ &{\rm Tr}(\rho_{ss}J^{+}_{k, l'}J^{-}_{k, m'}) \quad l \neq m , \ l' \neq m', \ l \neq m', \ l' \neq m.
\end{align}
We can easily choose the correct combination of indices on the permutation operators to lift the restrictions $l \neq m'$ and $l' \neq m$ and so we get:
\begin{equation}
\langle J^{+}_{k, l}J^{-}_{k, m} \rangle = \langle J^{+}_{k, l'}J^{-}_{k, m'} \rangle \quad l \neq m , \ l' \neq m'.
\label{Eq: Uniform Corr}
\end{equation}
Hence, we show that these correlations have no identifiable length scale as they are uniform for any of the symmetries $j$, which satisfy (\ref{Eq: Symmetries2}), if we can find a valid two-site permutation operator which commutes with all the operators  $\{O\} = \{J_{j}^{+-,z}\}\ \bigcup \ \{C\}$ at the same time. 
\par In the main text we choose spin dephasing $L_{k} = s_{k}^{z}$ to observe uniform, off-diagonal correlations in the $\eta$ SU(2) symmetry of the Hubbard model. This dephasing means the $\eta$ generators satisfy (\ref{Eq: Symmetries2}) but also that $S^{z}$ is an additional conserved quantity. Hence $\{O\} = \{\eta^{+},\eta^{-}, \eta^{z}, S^{z}\}$, and the kernel of $\Lambda$ is made up of the projectors of these operators (which are equivalent to the projectors of $\eta^{+}\eta^{-}$, $N_{\uparrow}$, $N_{\downarrow}$). We can find a valid permutation operator for all of these operators simultaneously - it must swap two sites and carry over a change of sign on the local operators when it does this. We consider initial states (e.g. ground states of the Hamiltonian) which typically have decaying short-range correlations in $\eta$ pairs. As $\langle \eta^{+}\eta^{-} \rangle$ is conserved these correlations will be spread over all length-scales of the system in order to satisfy Eq. (\ref{Eq: Uniform Corr}). Hence, in our simulations, we see uniform off-diagonal correlations in the $\eta$ sector.

\begin{figure*}[t]
\centering
\includegraphics[width = \textwidth]{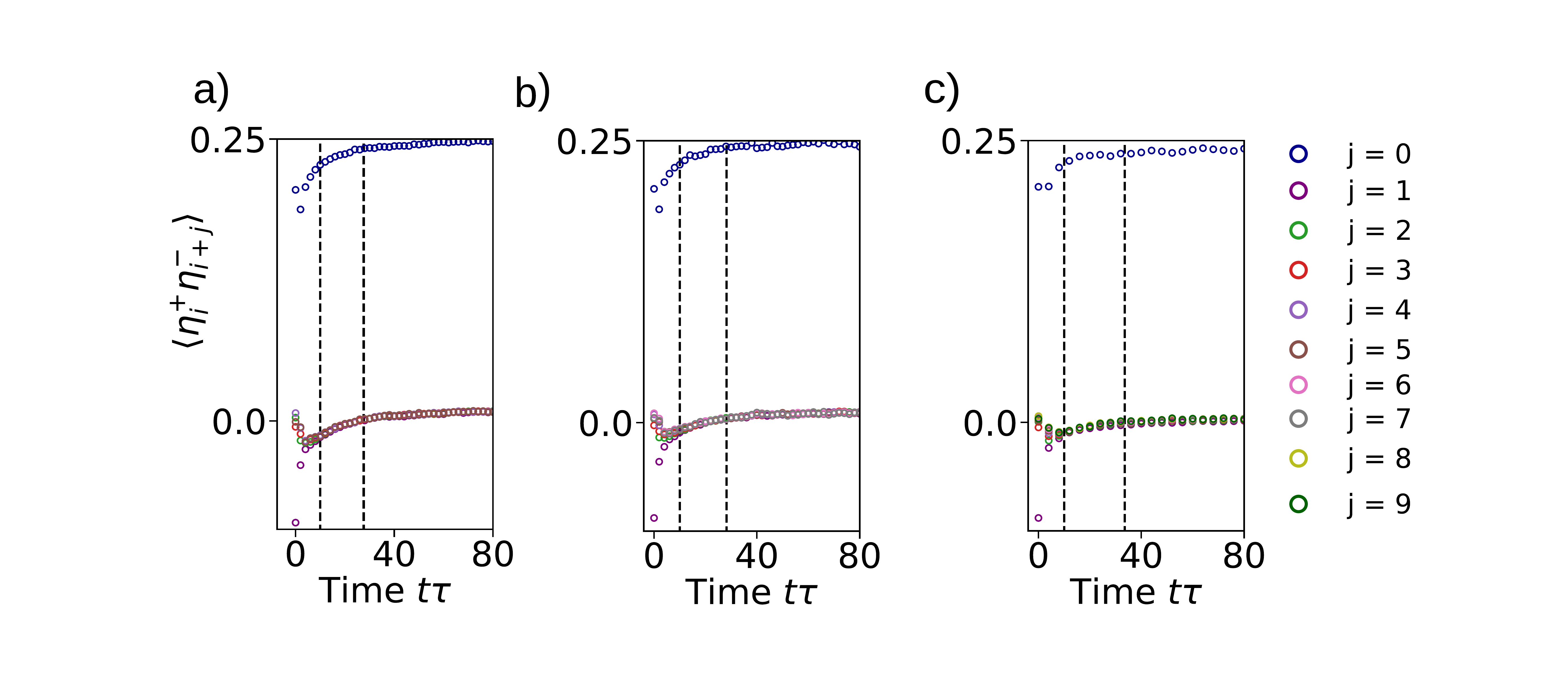}
\caption{(top) Dynamics of the, distance-averaged, quantity $\langle \eta^{+}_{i}\eta^{-}_{i + j} \rangle$ for a quench from the ground state of the $U = \tau$ Hubbard model at half-filling. During the quench dephasing is switched on with jump operators $s^{z}_{i}$ and $n_{i, \uparrow} + n_{i, \downarrow}$ on each site $i$ at rates $\gamma^{s} = 2.0\tau$ and $\gamma^{c} = 0.02\tau$ respectively. Dashed vertical lines are at $t\tau = 8.0$ and the time when the correlations have decayed to a $1/3$ of their value compared to those at $t\tau = 8.0$. (a) $L = 6$. (b) $L = 8$. (c) $L = 10$.}
\label{Fig:Perturbation}
\end{figure*}

\begin{figure}[t]
\centering
\includegraphics[width = \columnwidth]{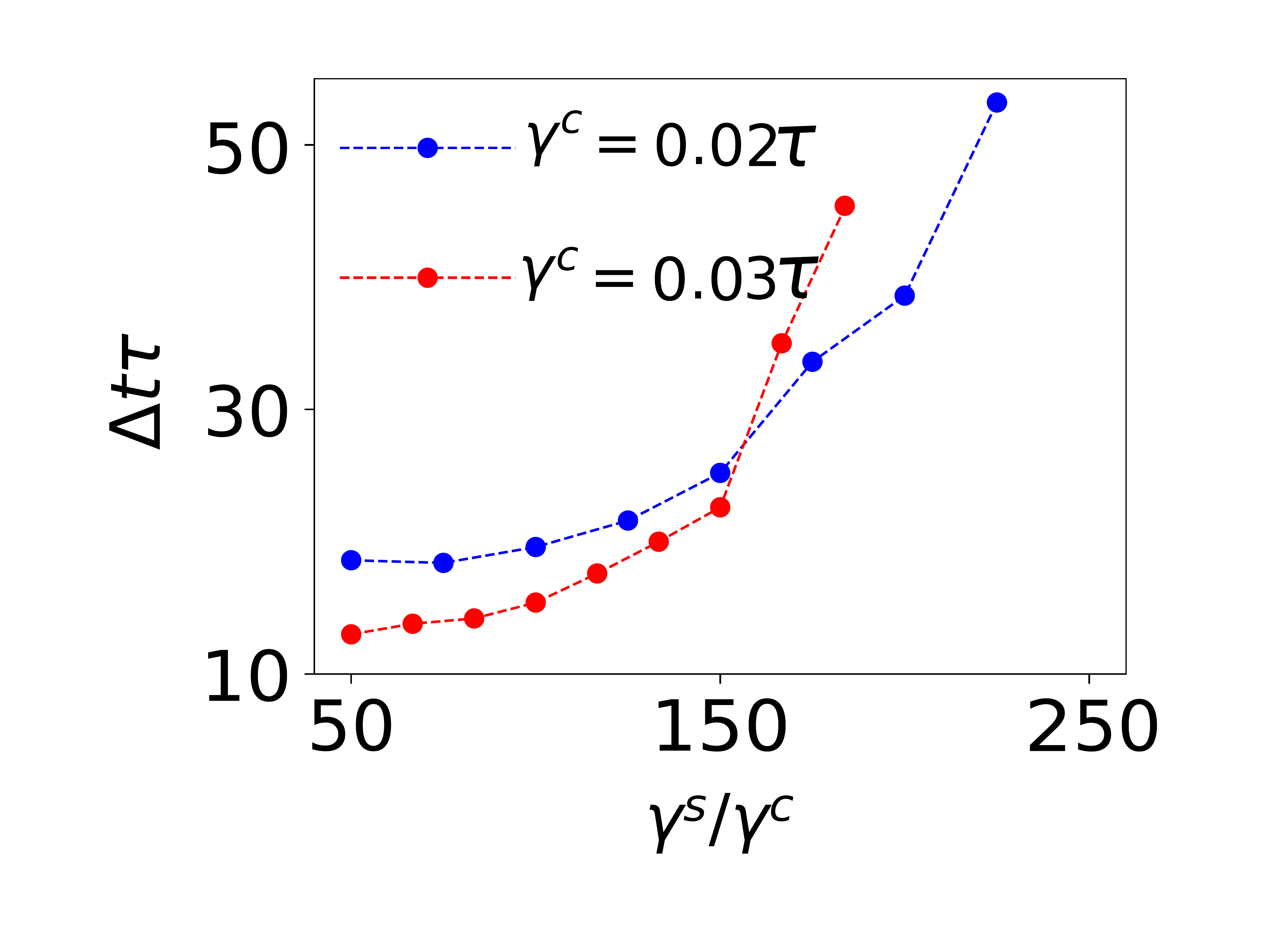}
\caption{Time $\Delta t \tau$ for $\eta$-correlations to decay to $1/3$ of their value at $t\tau = 8.0$ as a function of the ratio of the spin-dephasing to the charge dephasing for a quench starting from the ground state of the $L = 6, \ U = \tau$ Hubbard model. During the quench dephasing is applied with the values of of $\gamma_{c}$ and $\gamma_{s}$ specified in the plot.}
\label{Fig:PerturbationV2}
\end{figure}

\section{Stability of the master equation under perturbation}
We consider how the dynamics of our model is affected by perturbations in the dephasing. To simulate this we choose additional, `unwanted', dephasing in the number basis of the system - as well as having $\{L_{j}\} = s_{j}^{z}$ with constant rate $\gamma^{s} \ \ \forall j$ (see Eq. (\ref{Master EquationS})) we add in another set of local jump operators $\{L_{m}\} = n_{m, \uparrow} + n_{m, \downarrow}$ with constant rate $\gamma^{c}= \lambda \gamma^{s} \ \  \forall m$. The indices $j$ and $m$ each run over the lattice sites in the system. 
We have used local number dephasing because this additional dephasing breaks the strong symmetry relation $[\eta^{+}\eta^{-}, L_{m}] \neq 0 \ \forall m$. Consequently, in the long-time limit this additional dephasing completely destroys any coherences in the system and the steady state is a featureless infinite temperature ensemble. 
\par We set the amplitude of the unwanted dephasing to be around $2$ orders of magnitude less than the spin-dephasing. In cold atomic systems parameters can be tuned widely and unwanted decoherence mechanisms strongly suppressed \cite{Maciej1}. In quantum materials microscopic models are not as well established. However, the rates of different decoherence mechanisms can vary widely and careful engineering of these materials can significantly reduce their sensitivity to specific sources of decoherence \cite{QMat2}. As a result decoherence rates can easily differ by orders of magnitude, in line with the assumptions in our work. 
\par Figure \ref{Fig:Perturbation} shows that, for $\lambda = \gamma^{c} / \gamma^{s}=  0.01$, the additional dephasing does not prevent the observation of the ODLRO which forms and is maintained prior to the system decaying to a thermal classical ensemble. The window, indicated by the dashed lines, shows the time it takes for the off-diagonal correlations to decay to $1/3$ of their value at the first dashed line. The data in Fig. \ref{Fig:Perturbation} suggests the size of this window is not diminishing with system size. This is consistent with previous research studying the rate of relaxation in bulk-dephased many-body systems \cite{Gap1} or in random Liouvillians \cite{Gap2} - they find this rate either decays or saturates with system size.
\par Additionally, Fig. \ref{Fig:PerturbationV2} shows that for small perturbations this window can be lengthened by increasing the ratio $\gamma^{s} / \gamma^{c}$. Thus, in an experimental setup, being able to tune $\gamma_{s}$ could help mitigate the effects of the unwanted dephasing. Physically, the increased amplitude of the wanted (spin) dephasing compared to the unwanted (number) dephasing means the system is `measured’ with increasing frequency by the spin dephasing in comparison to the frequency with which it is `measured' by the number dephasing. As the ratio of these amplitudes increases then the spin dephasing effectively arrests the time evolution induced by the number-dephasing. Repeated measurements in the spin basis are freezing the system’s dynamics with respect to the dissipative evolution caused by the unwanted dephasing.

\begin{figure*}[t]
\includegraphics[width = \textwidth]{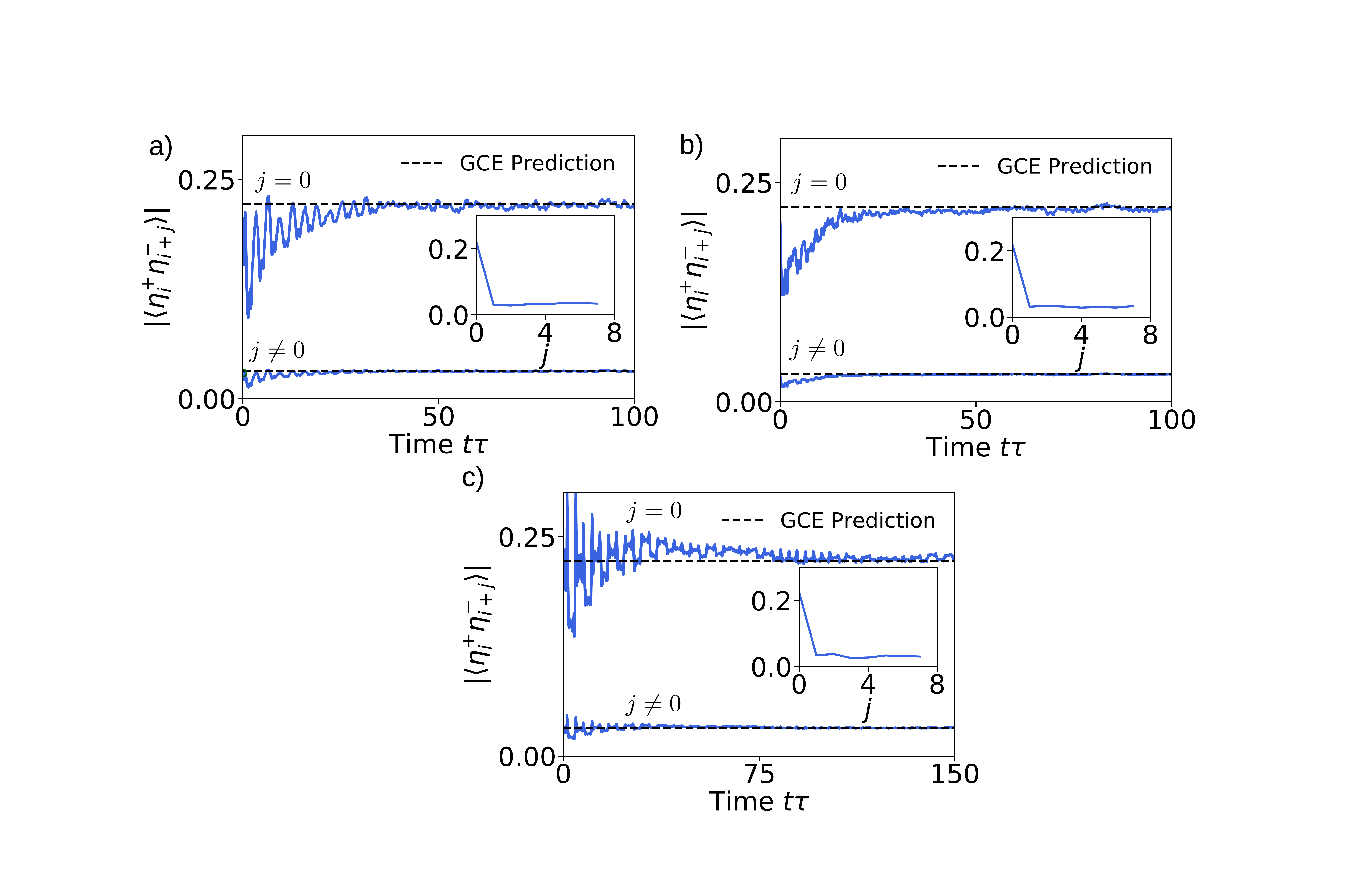}
\caption{Dynamics of the the $\eta$ correlations $|\langle \eta^{+}_{i}\eta^{-}_{i + j} \rangle|$ following a quench from the $U = \tau$ ground state of the $L = 8$ half-filled Hubbard model with $\langle S^{z} \rangle = 0$. The blue curves reflect the diagonal and off-diagonal correlations for a quench with the periodically driven Hamiltonian in Eq. (\ref{Eq:Driven Hubbard}) with $U = 6.0\tau$. The dashed lines indicate the prediction from the grand-canonical ensemble (GCE). (a) $V = 2.0U, \ \Omega = \tau$ and $f(i) = i$. (b) $V = 4.0U, \ \Omega = \tau$ and $f(i) = R[0, 2.0]$ where $R[0,2.0]$ is a random number in the range $0.0$ to $2.0$. (c) $V = 6.0U, \ \Omega = \tau$ and $f(i) = (-1)^{i}$. Insets) The dependence of the correlations on distance, extracted for the curves at time $t\tau = 100.0$.}
\label{Fig:Floquet}
\end{figure*}

\section{Floquet Heating}
In the main text we discussed that the properties of the steady states of (\ref{Master EquationS}) are reproduced by Floquet heating of the spin sector of the Hubbard model to infinite temperature. In order to show this we consider the Hubbard Hamiltonian with an additional time-dependent inhomogeneous magnetic field
\begin{align}
&H(t) = H_{\rm Hubbard}(U, \tau) + B(t)\sum_{i = 1}^{L}f(i)s^{z}_{i}, \notag \\ &\quad B(t) = V\cos(\Omega t),
\label{Eq:Driven Hubbard}
\end{align}
and $f(i)$ describes the inhomogeneity of the field. In Figure. \ref{Fig:Floquet} we plot the convergence of the $\eta$-correlations to those expected from the Grand Canonical Ensemble (GCE), $\rho_{\rm GCE} \propto \exp(\mu_{1}\eta^{+}\eta^{-} + \mu_{2}N_{\uparrow} + \mu_{3}N_{\downarrow})$, for $3$ different choices of inhomogeneity. As expected the choice of $f(i)$ is arbitrary and, along with the choice of system parameters, only affects the transient dynamics and the rate of relaxation. 

\section{Oscillating Coherences}
\par The Liouvillian in Eq. (\ref{Master EquationS}) also contains equally spaced imaginary eigenvalues \cite{DarkHamiltonians} due to the existence of a ladder operator, $\eta^{\pm}$, of the Hamiltonian which also commutes with the Lindblad operators. The associated eigenvectors take the form of a raising of the steady state $\rho_{ss}$: $\rho_{nm} \propto (\eta^{+})^{n}\rho_{ss}(\eta^{-})^{m}$ with the eigenvalues $\lambda_{nm} = 2 {\rm i}(m - n)\mu$. These states couple together sectors of the Hilbert Space with different particle numbers. 
\par If one were able to initialise the system in a state with coherent superpositions between states with different particle numbers then, in the long-time limit, the dynamics of observables such as $\eta^{x}$ should be oscillatory \cite{DarkHamiltonians}. In the main text we have focused solely on the properties of the steady state by enforcing a specific number of particles within our simulations.
\begin{figure*}[t]
\includegraphics[width = \textwidth]{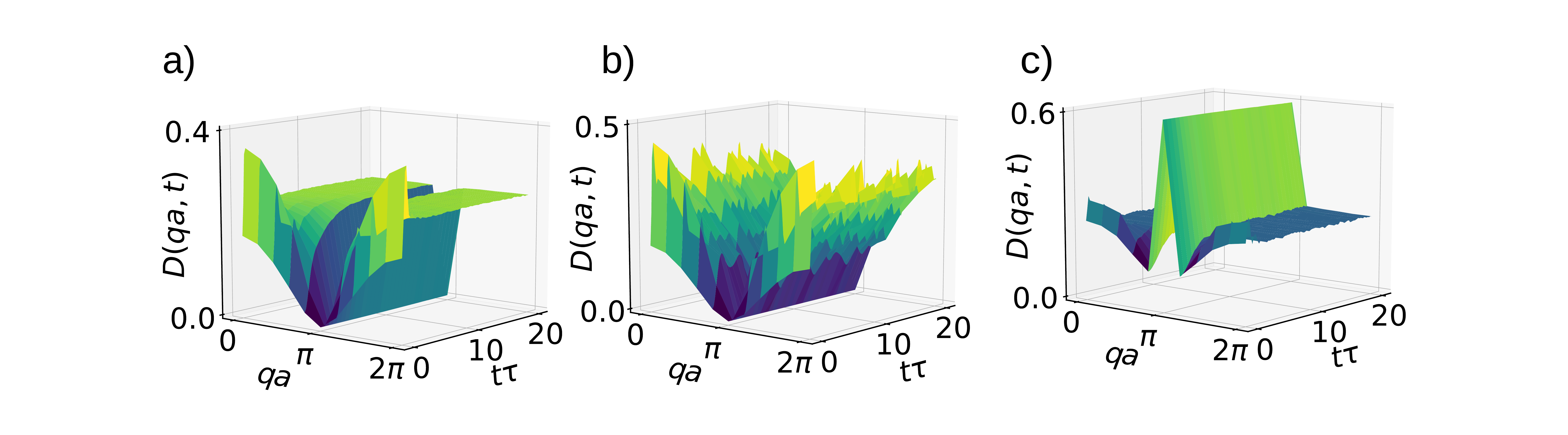}
\caption{Evolution of the structure factor $D(qa, t)$ in Eq. (\ref{Eq:D}) over time for the $10$ site Hubbard model at half-filling and $\langle S^{z} \rangle = 0$. The system is initialised in the ground state of $H$ for $U = 4.0 \tau$ and evolved under the same parameters but with a system-environment coupling of $\gamma$ and an interaction strength of $U = \tau$. (a) $\gamma = 2.0 \tau$. (b) $\gamma = 0.0$. (c) $\gamma = 2.0 \tau$ and the system is initialised in the state $ \propto (\eta^{+})^{2}\ket{\psi}$ where $\ket{\psi}$ is the ground state of the same Hamiltonian in a) and b) but with only $6$ total particles.}
\label{Fig:Doublons}
\end{figure*}

\section{Quasi-momentum Distributions}
\par We also plot the momentum distribution of the doublons following a quench from the ground state of the Hubbard model using the master equation (\ref{Master EquationS}). Figure \ref{Fig:Doublons}. shows the evolution of the doublon structure factor
\begin{align}
D(qa, t) = \frac{1}{L}\sum_{j, k = 1}^{L}\langle c^{\dagger}_{\uparrow, j}c^{\dagger}_{\downarrow, j}c_{\downarrow, k}c_{\uparrow, k}e^{{\rm i}(k-j)qa} \rangle(t),
\label{Eq:D}
\end{align} 
in time for the full range of quasi-momenta $q = 2\pi n / La$ \ $ n \in \{0, 1, ..., L - 1\}$, with $a$ the lattice spacing. In the long-time limit of the quench we see the flattening and equal excitation of all momenta modes, except for $qa  = \pi$. This is characteristic of the steady states of the map in Eq. (\ref{Master EquationS}) due to their off-diagonal long-range order. The occupation of the momentum mode at $qa  = \pi$ is equal to $\langle \eta^{+} \eta^{-} \rangle / L$, it is a constant of the evolution and for states 
with increasing values of $\langle \eta^{+}\eta^{-} \rangle$ the amplitude of this mode will grow (Fig. \ref{Fig:Doublons}c), indicating the presence of a doublon condensate \cite{Condensate}.

\section{Implementing Spin Dephasing in a Cold Atom Setup}
\par We now discuss the possibility of simulating the spin-dephased Hubbard model in a cold-atom setup. The Hamiltonian in Eq. \ref{HubbardHamS} can be accurately realised by loading an ultracold gas of fermionic atoms into a lattice structure created with counter-propagating laser beams \cite{Jaksch, Maciej, Maciej1}. These optical lattices provide the experimentalist precise control over the microscopic parameters of the system. 
\par In order to engineer dephasing solely in the spin degrees of freedom of the lattice we consider the possibility of immersing the lattice into a homogeneous Bose-Einstein Consendate (BEC), which was also discussed in \cite{DarkHamiltonians} in the context of achieving dephasing in the number basis of the Hubbard model. The interactions between the BEC and the lattice atoms create deformations in the condensate which can be described as polarons, or coherent states of phonons \cite{QME1}. 
\par To explicitly derive the dynamics of the lattice in this environment we closely follow methodology of \cite{QME1} and \cite{QME2}. We start by writing the Hamiltonian $H_{\rm tot} = H_{L} + H_{I} + H_{\rm BEC}$ of the total system
\begin{align}
&H_{L} = H, \notag \\
&H_{I} = \int \big(\kappa_{\uparrow}\chi_{\uparrow}^{\dagger}(\mathbf{r})\chi_{\uparrow}(\mathbf{r})\phi^{\dagger}(\mathbf{r})\phi(\mathbf{r}) +\hdots \notag \\ &\hdots \kappa_{\downarrow}\chi_{\downarrow}^{\dagger}(\mathbf{r})\chi_{\downarrow}(\mathbf{r})\phi^{\dagger}(\mathbf{r}) \phi(\mathbf{r}) \big) d \mathbf{r}, \notag \\
&H_{\rm BEC} = \int \phi^{\dagger}(\mathbf{r}) \bigg[- \frac{ \nabla^{2}}{2m_{b}} + V_{\rm ext}(\mathbf{r}) + \frac{g}{2}\phi^{\dagger}(\mathbf{r})\phi(\mathbf{r})   \bigg]\phi(\mathbf{r}) d \mathbf{r},
\end{align}
with $H_{L}, H_{I}$ and $H_{\rm BEC}$ being the lattice (see Eq. \ref{HubbardHamS}), interaction and BEC Hamiltonians respectively. The operators $\chi_{\uparrow}$ and $\chi_{\downarrow}$ are the field operators of the two fermionic levels which couple to the bosonic field ($\phi^{\dagger}(\mathbf{r})$) with amplitudes $\kappa_{\uparrow}$ and $\kappa_{\downarrow}$ respectively. The mass of a condensate atom is given by $m_{b}$, $V_{\rm ext}$ is an external trapping potential for the BEC and $g$ is the interaction strength between the condensate atoms.
\par In order for the condensate to affect the dynamics of the lattice solely in the spin sector we set the coupling strengths to be equal and opposite, i.e. $\kappa_{\uparrow} = - \kappa_{\downarrow} = \kappa$. This could be achieved via Feshbach resonances \cite{FeshbachOpt, FeshbachMag} which can be used to tune atomic scattering lengths over a wide range of values. It is also necessary that $\kappa$ is sufficiently small $\kappa \ll gn^{0}\epsilon^{D}$, where $n_{0}$ is the BEC density in the centre of the trap, $\epsilon$ is the healing length and $D$ is the system dimension \cite{QME1}. By solving the Gross-Pitaevskii equation and treating the resulting deformations in the BEC as coherent states of phonons an effective, discretized, Hamiltonian for the system can be derived. We then trace out the phonon degrees of freedom, invoking the rotating wave and Born approximations, to derive a Master equation for the density matrix of the lattice $\rho_{L}(t)$
\begin{align}
&\partial_{t}\rho_{L}(t) = -i[H_{g}, \rho_{L}(t)] + \hdots \notag \\ &\hdots \sum_{l, l' = 1}^{M}f_{l, l'}(t)\big(s^{z}_{l}s^{z}_{l'}\rho_{L}(t) + \rho_{L}(t)s^{z}_{l}s^{z}_{l'} -2s^{z}_{l}\rho_{L}(t)s^{z}_{l'}\big), \notag \\
&H_{g} = H^{'} - \sum_{l, l' = 1}^{M}g_{l, l'}(t)s^{z}_{l}s^{z}_{l'},
\label{Eq:ME}
\end{align}
where $H^{'}$ is the Hubbard Hamiltonian (the parameters have been slightly modified due to the BEC presence) and $f_{l, l'}(t)$ and $g_{l, l'}(t)$ are a pair of time-dependent, short-range (based on `typical' parameters for a BEC) functions which become time-independent in the long-time limit. Notably, however, the explicit form of these functions does not alter the strong symmetries of the system (they are still $\eta^{+}\eta^{-}$, $S^{z}$ and $N_{\rm total}$). Therefore the steady states of this Master equation are guaranteed to have long-range $\eta$-paired correlations in the long-time limit.

\end{document}